\documentclass[journal]{IEEEtran}
\usepackage{cite}
\usepackage{xcolor}
\usepackage{graphicx}
\usepackage{amssymb,amsmath,bm}
\usepackage[colorlinks,linkcolor=black,anchorcolor=black,citecolor=black,urlcolor=black]{hyperref}
\usepackage{hyperref}
\usepackage{multirow}
\usepackage{amsfonts}
\usepackage{url}
\usepackage{multirow}
\usepackage{color}
\usepackage{tabularx}
\usepackage{color}
\usepackage{diagbox}
\usepackage{bbding}
\usepackage{comment}
\usepackage{enumitem}
\usepackage{float}
\usepackage{booktabs}
\usepackage{threeparttable}
\ifCLASSINFOpdf
\else
\fi
\begin{document}
%
\title{LatentFlowSR: High-Fidelity Audio Super-Resolution via Noise-Robust Latent Flow Matching}
%
%
%

\author{Fei Liu, Yang~Ai,~\IEEEmembership{Member,~IEEE}, Hui-Peng Du, Yu-Fei Shi, ~Zhen-Hua~Ling,~\IEEEmembership{Senior Member,~IEEE}
\thanks{This work was funded by the National Nature Science Foundation of China under Grant 62301521.}
\thanks{F. Liu, Y. Ai, H.-P. Du, Y.-F. Shi, and Z.-H. Ling are with the National Engineering Research Center of Speech and Language Information Processing, University of Science and Technology of China, Hefei, 230027, China (e-mail: fliu215@mail.ustc.edu.cn, yangai@ustc.edu.cn, redmist@mail.ustc.edu.cn, zkddsr2023@mail.ustc.edu.cn, zhling@ustc.edu.cn).}
\thanks{Corresponding author: Yang Ai.}
}
%
%

\markboth{}%
{Shell \MakeLowercase{\textit{et al.}}: Bare Demo of IEEEtran.cls for Journals}
%



\maketitle

\begin{abstract}
Audio super-resolution aims to recover missing high-frequency details from bandwidth-limited low-resolution audio, thereby improving the naturalness and perceptual quality of the reconstructed signal. 
However, most existing methods directly operate in the waveform or time–frequency domain, which not only involves high-dimensional generation spaces but is also largely limited to speech tasks, leaving substantial room for improvement on more complex audio types such as sound effects and music.
To mitigate these limitations, we introduce LatentFlowSR, a new audio super-resolution approach that leverages conditional flow matching (CFM) within a latent representation space.
Specifically, we first train a noise-robust autoencoder, which encodes low-resolution audio into a continuous latent space.
Conditioned on the low-resolution latent representation, a CFM mechanism progressively generates the corresponding high-resolution latent representation from a Gaussian prior with a one-step ordinary differential equation (ODE) solver.
The resulting high-resolution latent representation is then decoded by the pretrained autoencoder to reconstruct the high-resolution audio.
Experimental results demonstrate that LatentFlowSR achieves competitive or superior performance compared with baseline methods across various audio types and super-resolution settings. 
These results indicate that the proposed method possesses strong high-frequency reconstruction capability and robust generalization performance, providing compelling evidence for the effectiveness of latent-space modeling in audio super-resolution.
\end{abstract}

\begin{IEEEkeywords}
audio super-resolution, conditional flow matching, autoencoder, latent representation
\end{IEEEkeywords}

%
\IEEEpeerreviewmaketitle

\section{Introduction}

\IEEEPARstart{I}
n real-world audio transmission scenarios, high-frequency components of audio signals are often lost due to limitations in transmission devices or channel bandwidth, which can negatively affect both perceptual quality and intelligibility. 
Therefore, audio super-resolution aims to reconstruct high-resolution audio signals from low-resolution inputs by restoring the missing high-frequency information, thereby improving the naturalness and intelligibility of the audio.
High-frequency reconstruction capability enables audio super-resolution to be applied to various downstream tasks, such as text-to-speech (TTS) synthesis \cite{nakamura2014mel}, automatic speech recognition (ASR) \cite{goodarzi2012feature,albahri2016artificial}, speaker recognition \cite{li2019speech,li2015dnn,nidadavolu2018investigation}, audio coding \cite{xiao2023multi} and music restoration \cite{moliner2022behm}. 
By restoring missing high-frequency details, it compensates for information loss, thereby improving synthesis quality, reducing word error rates in recognition tasks, and enhancing performance across various downstream applications.

Early audio super-resolution techniques were mainly based on traditional signal processing methods, including interpolation-based approaches, parametric speech-based approaches, and mapping-based approaches, with most of these early methods primarily designed for speech signals \cite{prasad2016bandwidth,jax2003artificial,chennoukh2001speech,uysal2005bandwidth,abe1994algorithm,nakatoh1997generation,park2000narrowband}.
Interpolation-based methods \cite{prasad2016bandwidth,jax2003artificial} typically perform interpolation in the time domain and then apply a low-pass filter to remove the imaging components introduced during interpolation. 
Parametric speech-based methods \cite{chennoukh2001speech,uysal2005bandwidth} model speech signals using an excitation–envelope representation, where the low-frequency spectral envelope is extended to the high-frequency range and high-frequency excitation is generated to reconstruct the high-frequency speech components. 
Mapping-based methods \cite{abe1994algorithm,nakatoh1997generation,park2000narrowband} map low-frequency speech features to high-frequency features through techniques such as codebook mapping or linear mapping, and then reconstruct the corresponding high-frequency speech signals.
Although these signal-processing-based methods are relatively simple and highly interpretable, they often struggle to reconstruct natural high-frequency details. 
As a result, the generated speech tends to sound muffled, and these methods generally exhibit limited generalization ability.

With the rapid advancement of deep learning, neural-network-based audio super-resolution methods have achieved impressive results, but the early research in this field was largely centered on speech super-resolution.
These approaches can generally be categorized into waveform-based methods and spectrum-based methods.
Waveform-based methods directly map low-resolution waveforms to high-resolution waveforms in the time domain using neural networks \cite{kuleshov2017audio,wang2021towards,zhang2021wsrglow,lee2021nu,han2022nu}. For example, NU-Wave \cite{lee2021nu} was the first to introduce diffusion probabilistic models into the field of audio super-resolution. Conditioned on the low-resolution audio signal, it leverages the strong distribution modeling capability of diffusion models to recover missing high-frequency details and directly generate waveforms in the time domain. NU-Wave2 \cite{han2022nu} further improves upon NU-Wave, achieving better efficiency and generation quality. However, waveform-based approaches require processing each time-domain sample individually, which leads to low generation efficiency. This limitation becomes particularly significant for long high-resolution audio sequences, making such methods less practical in real-world applications \cite{liu2022neural}.

Spectrum-based methods, on the other hand, perform modeling in the frequency domain and use neural networks to predict the spectral features of high-resolution waveform \cite{mandel2023aero,shuai2023mdctgan,lu2024towards,yu2023conditioning}. mdctGAN \cite{shuai2023mdctgan} models the modified discrete cosine transform (MDCT) spectrum and incorporates adversarial learning to predict a high-resolution MDCT spectrum from the low-resolution input, followed by an inverse transform to reconstruct the high-resolution audio. AP-BWE \cite{lu2024towards} explicitly models the amplitude and phase spectra, predicting both components in parallel before reconstructing the audio through an inverse transform. UDM+ \cite{yu2023conditioning} further formulates audio super-resolution as a spectral inpainting problem: during the reverse diffusion sampling process, the low-frequency components of the spectrum are directly replaced, and only the missing high-frequency information is generated.
Although these methods achieve promising high-frequency reconstruction performance, they primarily focus on speech super-resolution, with limited exploration of sound effects, music, etc., resulting in a relatively narrow range of applications.

Recently, some audio super-resolution studies have begun to focus on audio signals beyond speech \cite{liu2024audiosr,im2025flashsr,yun2025flowhigh,li2025audio}. 
Other types of audio, e.g., sound effects and music, differ significantly from speech in structure. They exhibit fundamentally different harmonic patterns, with greater diversity and irregularity, and their high-frequency components contain abundant timbral details and texture-related information \cite{siedenburg2019timbre,schubert2004spectral}. These components are more difficult to recover than speech high-frequency content, thus requiring more powerful super-resolution models.
For example, AudioSR \cite{liu2024audiosr} employs a latent diffusion model (LDM) to model the mel-spectrogram, recovering a high-resolution mel-spectrogram from a low-resolution one and then reconstructing high-resolution audio using a vocoder. FlashSR \cite{im2025flashsr}, built upon the LDM framework of AudioSR, distills the model into a single-step student LDM that restores high-resolution mel-spectrograms, significantly improving inference speed while maintaining quality close to that of AudioSR. 
FlowHigh \cite{yun2025flowhigh} introduces conditional flow matching (CFM) into audio super-resolution task, learning a velocity field in the mel-spectrogram domain and sampling from it to generate a high-resolution mel-spectrogram, which is then converted back to high-resolution audio using a vocoder.
Although these approaches can be applied to a wider variety of audio signals, most of them are still limited to modeling in the mel-spectrogram space, where phase information is discarded and must later be reconstructed by a vocoder. 
A few studies have explored discrete latent spaces, but discretization may cause information loss \cite{li2025audio}.
As a result, existing methods still leave substantial room for improvement in terms of high-resolution audio reconstruction quality.

\begin{figure}
    \centering
    \includegraphics[width=\linewidth]{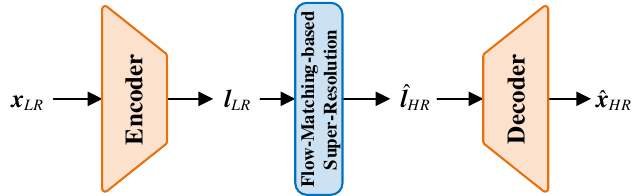}
    \caption{Overview of the proposed LatentFlowSR.}
    \label{fig:p1}
\end{figure}

Therefore, we propose a new audio super-resolution model, named LatentFlowSR, which performs flow matching \cite{lipman2023flow} in the latent space. Specifically, we first train a noise-robust autoencoder and use it to map low-resolution audio into a compressed low-resolution latent representation.
Next, in the latent space, we employ a CFM mechanism to perform super-resolution. 
Conditioned on the low-resolution latent, the CFM mechanism starts from Gaussian noise and predicts a velocity field to guide the generative process. 
By approximately integrating this velocity field with ordinary differential equation (ODE) solver, a high-resolution latent representation is obtained, which is finally decoded by the pretrained autoencoder’s decoder to reconstruct the high-resolution audio.
Experimental results demonstrate that LatentFlowSR achieves state-of-the-art performance across multiple super-resolution tasks on diverse audio types compared with methods that directly model signals in either the time or frequency domain, thereby verifying the effectiveness of super-resolution modeling in the latent space. 
Furthermore, the experimental results suggest that latent-space modeling can ease the super-resolution process, thereby promoting improvements in efficiency and computational complexity. 

This paper is organized as follows.
In Section \ref{sec:propose}, we provide the details of our proposed LatentFlowSR model. 
In Section \ref{sec:exp}, we describe the specific details of the experimental setup.
In Section \ref{sec:res}, we present our experimental results.
Finally, we give conclusions in Section \ref{sec:con}.

\begin{figure*}
    \centering
    \includegraphics[width=\linewidth]{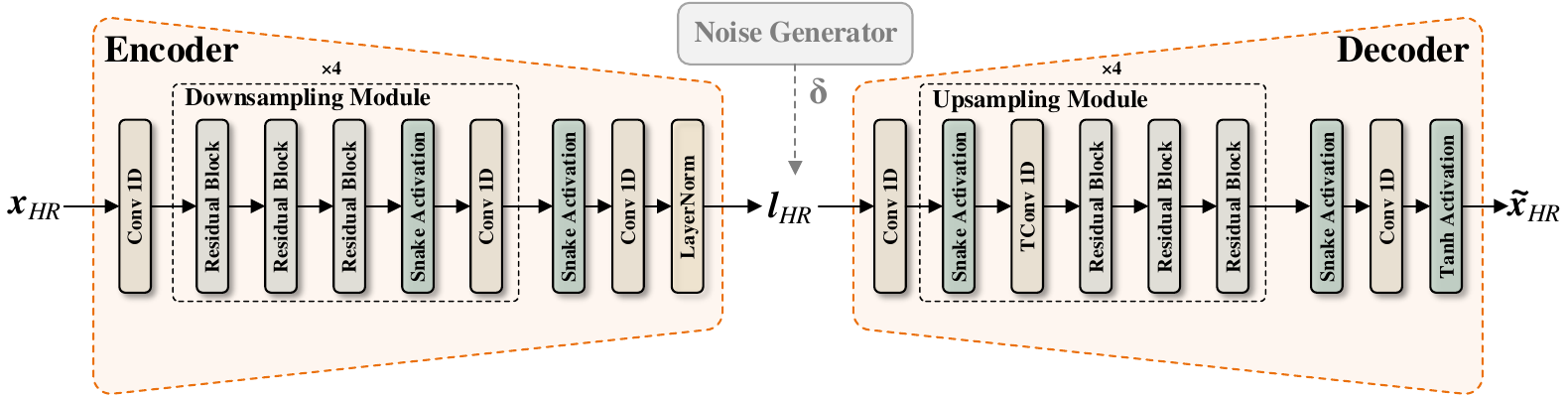}
    \caption{Overview of the noise-robust autoencoder. The noise generator only appears during the training process.}
    \label{fig:p2}
\end{figure*}

\section{Proposed Method}
\label{sec:propose}

\subsection{Overview}

In this paper, we propose LatentFlowSR, a CFM–based audio super-resolution method that performs modeling in the latent space. 
LatentFlowSR mainly consists of a noise-robust autoencoder and a CFM-based super-resolution model. 
The overall framework is illustrated in Figure \ref{fig:p1}.
Given a low-resolution audio waveform $\bm{x}_{LR}'\in \mathbb{R}^{L'}$, we first upsample it to the target resolution to obtain $\bm{x}_{LR}\in \mathbb{R}^{L}$ which still lacks the high-frequency bands, where $L'$ and $L$ denote the waveform lengths of $\bm{x}_{LR}'$ and $\bm{x}_{LR}$, respectively, and thus the super-resolution ratio is $L/L'$. 
The upsampled audio is then compressed by the encoder of the autoencoder into a low-resolution latent representation $\bm{l}_{LR}\in \mathbb{R}^{C \times T}$, where $C$ and $T$ denote the number of channels and frames, respectively.
This representation is then fed into the super-resolution model to recover a high-resolution latent representation $\bm{\hat{l}}_{HR}\in \mathbb{R}^{C \times T}$.
Finally, the decoder of the autoencoder reconstructs the high-resolution audio $\bm{\hat{x}}_{HR}\in \mathbb{R}^{L}$ from $\bm{\hat{l}}_{HR}$.
To fully exploit the information in the low-resolution audio, we replace the low-frequency band of the generated audio signal with the corresponding low-frequency band from the input audio \cite{yun2025flowhigh,yu2023conditioning,liu2024audiosr,im2025flashsr,liu2022neural,li2025audio}.
In the following sub-sections, we describe each component in detail.

\subsection{Noise-Robust Autoencoder}

The noise-robust autoencoder mainly consists of an encoder and a decoder, and its architecture is illustrated in Figure \ref{fig:p2}.
During training, the noise-robust autoencoder jointly trains the encoder and decoder using high-resolution data.
The original high-resolution audio $\bm{x}_{HR}\in \mathbb{R}^{L}$ is first processed by the encoder and compressed into a low-dimensional latent space, producing the latent representation $\bm{l}_{HR}\in \mathbb{R}^{C \times T}$.
This latent representation is then passed through the decoder to reconstruct the high-resolution audio $\bm{\tilde{x}}_{HR}\in \mathbb{R}^{L}$.
In the subsequent super-resolution process, the encoder and decoder are respectively used for low-resolution latent representation extraction and high-resolution audio reconstruction.
The detailed design of the noise-robust autoencoder is described as follows.

\subsubsection{Encoder}
The encoder is composed of multiple one-dimensional convolution and downsampling modules. 
The input $\bm{x}_{HR}$ is first processed by a one-dimensional convolution layer to extract temporal features. 
It then passes through four downsampling modules that progressively reduce the temporal resolution while increasing the channel dimension, thereby shortening the feature representation and reducing computational cost for latent-space super-resolution while preserving essential information \cite{zeghidour2021soundstream,defossezhigh,kumar2023high}.
Each downsampling module consists of three residual blocks, a learnable periodic activation function, i.e., Snake, and a one-dimensional convolution layer for downsampling. 
Within each residual block, two Snake activations and two one-dimensional convolution layers are stacked alternately, and a residual connection \cite{he2016deep,he2016identity} is employed to add the input of the block to its output. 
The Snake activation introduces a learnable periodic component, enabling the network to better model the periodic structures commonly found in audio signals \cite{ziyin2020neural,lee2023bigvgan}. 
After downsampling, the features are further fused across channels through an additional combination of a Snake activation and a one-dimensional convolution layer. Finally, layer normalization is applied to stabilize the channel scale and produce a compact latent representation $\bm{l}_{HR}$.

\subsubsection{Decoder}
The decoder adopts a structure that is approximately mirror-symmetric to the encoder. The compressed latent representation $\bm{l}_{HR}$ first passes through a one-dimensional convolution layer to increase the channel dimension, and then enters four upsampling modules that progressively restore the temporal resolution.
In each upsampling module, the features are first activated by a Snake function and then upsampled using a one-dimensional transposed convolution. This is followed by residual blocks that refine the upsampled fine-grained structures. Finally, the feature representation is further processed by an additional Snake activation and a one-dimensional convolution layer, and a Tanh activation is applied to constrain the output amplitude range, producing the reconstructed high-resolution waveform $\bm{\tilde{x}}_{HR}$ in the time domain.

\begin{figure*}
    \centering
    \includegraphics[width=\linewidth]{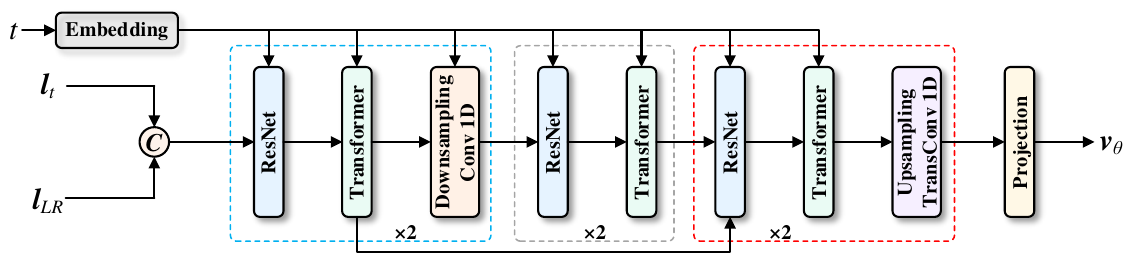}
    \caption{Overview of the velocity field estimation network used in CFM mechanism.}
    \label{fig:p3}
\end{figure*}

\subsubsection{Noise-Robust Training Strategy}

During autoencoder training, the decoder takes the target latent representation as input. 
In contrast, during the inference stage of LatentFlowSR, the decoder is fed with the predicted latent representation, and this train–inference mismatch may degrade its decoding performance. 
To enable the decoder to adapt in advance to perturbed latent representations and thus improve its robustness, we sample $\delta\sim \mathcal{N}(\bm{0}, \bm{I})$ from a standard Gaussian distribution and add it into the compressed latent representation $\bm{l}_{HR}$ during autoencoder training, where $\bm{I}$ is a $C \times T$-dimensional identity covariance matrix.

Meanwhile, the autoencoder is optimized using both adversarial loss and waveform reconstruction loss. 
We employ a composite discriminator adopted from \cite{kumar2023high}.
The generator (i.e., the autoencoder) and discriminator losses of the adversarial training are defined as follows:
\begin{equation}
    \mathcal L_{G}=\mathbb{E}_{\bm{\tilde{x}}_{HR}}\text{max}(0,1-D(\bm{\tilde{x}}_{HR})),
\end{equation}
\begin{equation}
    \mathcal L_{D}=\mathbb{E}_{(\bm{x}_{HR},\bm{\tilde{x}}_{HR})}[\text{max}(0,1-D(\bm{x}_{HR}))+\text{max}(0,1+D(\bm{\tilde{x}}_{HR}))], 
\end{equation}
where $D$ denotes the discriminator. 
The waveform reconstruction loss is formulated as the L1 distance between the input audio waveform $\bm{x}_{HR}$ and the reconstructed audio waveform $\bm{\tilde{x}}_{HR}$, i.e.,
\begin{equation}
    \mathcal L_{R}=\mathbb{E}_{(\bm{x}_{HR},\bm{\tilde{x}}_{HR})}\left\|  \bm{x}_{HR}-\bm{\tilde{x}}_{HR} \right\|_{1}, 
\end{equation}
where $\left\| \cdot \right\|_{1}$ is the L1 norm.
We use $\mathcal{L}_G+\mathcal{L}_R$ and $\mathcal{L}_D$ to train the generator and discriminator alternately, following the standard adversarial training paradigm.

\subsection{CFM-based Latent-Space Super-Resolution}
We employ a CFM mechanism $\phi_{CFMSR}$ \cite{tong2024improving,liu2023flow} to perform super-resolution in the latent space. 
The mechanism takes the low-resolution latent representation $\bm{l}_{LR}$, which is obtained by encoding the upsampled audio $\bm{x}_{LR}$ with the pretrained autoencoder's encoder, as conditional information and learns a time-dependent velocity field to map an initial distribution $\bm{l}_0 \sim \mathcal{N}(\bm{0}, \bm{I})$ to the desired high-resolution latent representation $\bm{\hat{l}}_{HR}\in \mathbb{R}^{C \times T}$, i.e.,
\begin{equation}
    \bm{\hat{l}}_{HR}=\phi_{CFMSR}(\bm{l}_0|\bm{l}_{LR}).
\end{equation}
Finally, $\bm{\hat{l}}_{HR}$ is passed through the decoder of the pretrained autoencoder to reconstruct the high-resolution audio $\bm{\hat{x}}_{HR}$.

\subsubsection{CFM Mechanism for Latent-Space Modeling}

We define a CFM time-dependent state $\bm{l}_t\in \mathbb{R}^{C \times T}$ in the latent space, where $t\in[0,1]$.
The state at $t=0$ corresponds to the initial state $\bm{l}_0$, while the state at $t=1$ corresponds to the high-resolution latent representation $\bm{l}_{HR}$ (i.e., $\bm{l}_{1}=\bm{l}_{HR}$).
The evolution of this process is described by an ODE, i.e.,
\begin{equation}
    \frac{d\bm{l}_t}{dt}=\bm{v}_{\theta}(\bm{l}_t,t,\bm{l}_{LR}),
\end{equation}
where $\bm{v}_{\theta}$ denotes the velocity field parameterized by neural network parameters $\theta$.
The velocity field takes the current state $\bm{l}_t$, the time variable $t$, and the conditional latent representation $\bm{l}_{LR}$ as inputs, and models the instantaneous direction that drives samples to evolve from the Gaussian prior toward the conditional target distribution $p(\bm{l}_{1}|\bm{l}_{LR})$.
The terminal state $\bm{l}_{1}$ is obtained from the initial state $\bm{l}_0$ by solving the following integral equation:
\begin{equation}
\label{eq:6}
    \bm{l}_{1}=\bm{l}_0+\int_{0}^{1} \bm{v}_{\theta}(\bm{l}_t,t,\bm{l}_{LR})dt.
\end{equation}
By conditioning on the low-resolution latent representation, the generation of the high-resolution latent is guided by stable semantic and acoustic information, enabling more effective reconstruction of high-frequency details.

\subsubsection{Velocity Field Estimation Network}
To estimate the velocity field $\bm{v}_{\theta}(\bm{l}_t,t,\bm{l}_{LR})$, we employ a U-Net–based generative network that combines local and global modeling \cite{ronneberger2015u}, as illustrated in Figure \ref{fig:p3}. 
The network takes the intermediate state $\bm{l}_t$ and the low-resolution latent representation $\bm{l}_{LR}$ as inputs and predicts the velocity field under the modulation of a time embedding from time $t$.

Specifically, at the input end, the intermediate state $\bm{l}_t$ and the low-resolution latent representation $\bm{l}_{LR}$ are concatenated along the channel dimension and fed into the subsequent network for modeling. 
This design allows the conditional information to provide semantic and acoustic guidance throughout the generative process, thereby improving the consistency and stability of the generated results \cite{rombach2022high,wang2023audit}. 
Meanwhile, to capture the dynamic characteristics of the continuous flow at different time steps, the time variable $t$ is mapped into a high-dimensional representation through a time-embedding module. This embedding is injected into network blocks, enabling the network to explicitly perceive the current time step and adjust the modeling of the instantaneous velocity accordingly \cite{chen2024binarized}.

The backbone follows a U-Net-style architecture and mainly consists of two downsampling blocks, two feature processing blocks, and two upsampling blocks.
1) Each downsampling block is composed of a ResNet block \cite{he2016deep}, a Transformer block \cite{vaswani2017attention}, and a one-dimensional downsampling convolution layer. 
Here, the ResNet block adopts a one-dimensional convolution-based residual structure consisting of two convolution layers, each followed by a group normalization layer \cite{dhariwal2021diffusion} and a Mish activation function \cite{min2021meta}. 
After the first convolution, the time embedding is added to the intermediate features of the main branch. Meanwhile, a residual branch is preserved, where the input features pass through a one-dimensional convolution and are added to the output of the main branch. This residual design mitigates gradient degradation during deep network training and improves the stability of local feature modeling \cite{he2016deep,he2016identity}.
Built upon the local features extracted by the ResNet block, the Transformer block further performs global modeling to capture long-range contextual dependencies. The downsampling convolution layer then reduces the temporal resolution of the features while enlarging the effective receptive field.
2) The feature processing blocks are placed in the middle of the network and consist of alternating ResNet and Transformer blocks, where the same ResNet design is used to refine local temporal patterns and the Transformer blocks further enhance global dependency modeling. 
These blocks continuously integrate local acoustic details with global temporal dependencies to refine the intermediate representations \cite{vaswani2017attention}.
3) Each upsampling block is composed of a ResNet block, a Transformer block, and an upsampling transposed convolution layer. In these blocks, the ResNet and Transformer blocks further enhance the intermediate representations, while the transposed convolution layer progressively restores the temporal resolution and reconstructs more detailed temporal structures. 
In addition, U-Net-style skip connections are introduced between the corresponding downsampling and upsampling stages, allowing shallow features to be fused with deeper representations so as to better preserve fine-grained local temporal details while maintaining long-range temporal coherence.
Through this hierarchical design, the network gradually combines short-term acoustic structures with long-range temporal coherence at different levels, enabling the velocity field prediction to recover fine-grained and globally consistent acoustic representations \cite{rakotonirina2021self,stoller2018waveunet}.

\begin{figure}
    \centering
    \includegraphics[width=\linewidth]{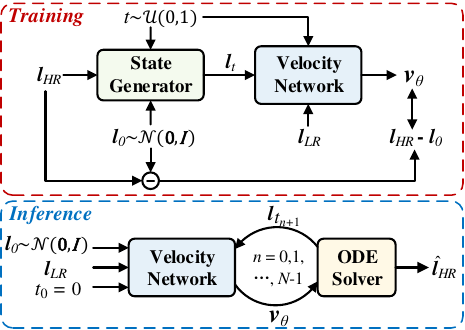}
    \caption{Overview of the CFM training and inference.}
    \label{fig:p4}
\end{figure}


At the output end, a projection block maps the features extracted by the backbone network to the velocity field.
This module mainly consists of a one-dimensional convolution, a group normalization layer, a Mish activation function, and a final one-dimensional convolution layer. 
It performs feature projection and channel mapping while preserving the temporal structure, ultimately producing the predicted velocity field $\bm{v}_{\theta}(\bm{l}_t,t,\bm{l}_{LR})$.

\subsubsection{Training and Inference}
Figure \ref{fig:p4} illustrates the training and inference processes of the CFM mechanism.
During CFM training, the autoencoder provides the target latent representation $\bm{l}_{HR}$ and the conditional input $\bm{l}_{LR}$. 
We then sample the initial state $\bm{l}_0$ from a standard Gaussian distribution $\mathcal{N}(\bm{0}, \bm{I})$ and uniformly sample the time variable $t\sim \mathcal{U}(0,1)$ from the interval $[0,1]$. To improve training efficiency and reduce optimization difficulty, we adopt the optimal transport CFM (OT-CFM) \cite{tong2024improving} formulation. Specifically, a linear path is constructed between the initial state $\bm{l}_0$ and the target state $\bm{l}_{HR}$, as defined below, 
\begin{equation}
    \bm{l}_t=(1-t)\bm{l}_0+t\bm{l}_{HR}.
\end{equation}
Based on this formulation, the corresponding target velocity field can be derived
\begin{equation}
    \bm{v}=\frac{d\bm{l}_t}{dt}=\bm{l}_{HR}-\bm{l}_0,
\end{equation}
which is independent of the time variable $t$.
The velocity field estimation network is then trained by minimizing the mean squared error between the predicted velocity field and the target one, i.e.,
\begin{equation}
    \mathcal{L}_{CFM}=\mathbb{E}_{\bm{l}_0 \sim \mathcal{N}(\bm{0}, \bm{I}), t\sim \mathcal{U}(0,1)}\left\| \bm{v}_{\theta}(\bm{l}_t,t,\bm{l}_{LR})-\bm{v} \right\|_2^2.
\end{equation}
In this way, OT-CFM only requires matching the instantaneous local velocity at randomly sampled intermediate states and time steps during training, which generally leads to more efficient and stable optimization.

During CFM inference, we take $\bm{l}_0$ as the initial state, and under the conditioning of $\bm{l}_{LR}$, use an Euler-based ODE solver to discretely approximate the integral in Equation \ref{eq:6}, thereby iteratively generating the predicted high-resolution latent representation $\bm{\hat{l}}_{HR}$.
Specifically, the time interval $[0,1]$ is uniformly divided into $N$ steps, and each discrete time point is defined as $t_n=n\Delta t$, where $\Delta t=\frac{1}{N}$ and $n=0,1,\cdots,N-1$.
Equation \ref{eq:6} can therefore be rewritten as
\begin{equation}
    \bm{l}_{t_{n+1}}=\bm{l}_{t_{n}}+\Delta t\cdot\bm{v}_{\theta}(\bm{l}_{t_{n}},t_n,\bm{l}_{LR}).
\end{equation}
After $N$ iteration steps, the final state $\bm{l}_{t_N}$ is regarded as the predicted high-resolution latent representation $\bm{\hat{l}}_{HR}$.

\begin{table*}[t]
    \centering
    \caption{Objective comparison of super-resolution performance of different methods on multiple datasets.}
    \label{tab:all_datasets}
    \small
    \setlength{\tabcolsep}{2.5pt}
    \resizebox{\textwidth}{!}{
    \begin{tabular}{cccccccccccccc}
        \toprule
        \multirow{2}{*}{Dataset} & \multirow{2}{*}{Method}
        & \multicolumn{3}{c}{8 kHz $\rightarrow$ 44.1 kHz}
        & \multicolumn{3}{c}{12 kHz $\rightarrow$ 44.1 kHz}
        & \multicolumn{3}{c}{16 kHz $\rightarrow$ 44.1 kHz}
        & \multicolumn{3}{c}{24 kHz $\rightarrow$ 44.1 kHz} \\
        \cmidrule(lr){3-5} \cmidrule(lr){6-8} \cmidrule(lr){9-11} \cmidrule(lr){12-14}
        & 
        & LSD $\downarrow$ & LSD-HF $\downarrow$ & ViSQOL $\uparrow$
        & LSD $\downarrow$ & LSD-HF $\downarrow$ & ViSQOL $\uparrow$
        & LSD $\downarrow$ & LSD-HF $\downarrow$ & ViSQOL $\uparrow$
        & LSD $\downarrow$ & LSD-HF $\downarrow$ & ViSQOL $\uparrow$ \\
        \midrule

        \multirow{9}{*}{\shortstack[c]{VCTK\\{(In-domain)}}}
        & NU-Wave      & -    & -    & -    & -    & -    & -    & 0.98 & 1.08 & 2.36 & 0.85 & 1.03 & 3.18 \\
        & NU-Wave2     & 1.10 & 1.19 & 2.48 & 0.93 & 1.08 & 2.75 & 0.86 & 1.03 & 3.00 & 0.73 & 0.97 & 3.74 \\
        & UDM+         & 1.03 & 1.13 & 2.48 & 0.89 & 1.02 & 3.08 & 0.79 & 0.96 & 3.30 & 0.65 & 0.88 & 4.00 \\
        & AudioSR      & 1.62 & 1.73 & 2.57 & 1.50 & 1.66 & 2.62 & 1.42 & 1.62 & 2.68 & 1.21 & 1.44 & 3.16 \\
        & FlashSR      & 1.71 & 1.85 & 2.41 & 1.45 & 1.64 & 2.57 & 1.36 & 1.62 & 2.64 & 0.96 & 1.23 & 2.82 \\
        & FlowHigh     & 1.29 & 1.32 & 3.34 & 1.25 & 1.29 & 3.43 & 1.23 & 1.28 & 3.58 & 1.17 & 1.23 & 4.08 \\
        & mdctGAN      & 0.93 & 1.01 & 3.03 & 0.85 & 1.02 & 3.12 & 0.83 & 0.98 & 3.27 & 0.71 & 0.96 & 3.69 \\
        & AP-BWE       & \textbf{0.86} & 0.97 & 3.34 & 0.79 & 0.92 & 3.42 & 0.74 & 0.90 & 3.60 & 0.63 & 0.87 & 4.15 \\
        & LatentFlowSR & \textbf{0.86} & \textbf{0.95} & \textbf{3.38} & \textbf{0.78} & \textbf{0.91} & \textbf{3.51} & \textbf{0.73} & \textbf{0.89} & \textbf{3.62} & \textbf{0.60} & \textbf{0.85} & \textbf{4.17} \\
        \midrule

        \multirow{4}{*}{\shortstack[c]{ESC-50\\{(Out-of-domain)}}}
        & AudioSR      & 1.75 & 1.88 & 2.78 & 1.69 & 1.89 & 2.83 & 1.65 & 1.93 & 2.84 & 1.58 & 2.07 & 3.11 \\
        & FlashSR      & 1.70 & 1.86 & 2.88 & 1.64 & 1.87 & 2.93 & 1.60 & 1.94 & 2.96 & 1.54 & 2.14 & 3.14 \\
        & FlowHigh     & 1.83 & 1.97 & 2.63 & 1.76 & 1.98 & 3.25 & 1.78 & 2.13 & 3.52 & 1.63 & 2.22 & 4.03 \\
        & LatentFlowSR & \textbf{1.36} & \textbf{1.50} & \textbf{3.19} & \textbf{1.30} & \textbf{1.51} & \textbf{3.34} & \textbf{1.31} & \textbf{1.62} & \textbf{3.57} & \textbf{1.18} & \textbf{1.71} & \textbf{4.11} \\
        \midrule

        \multirow{4}{*}{\shortstack[c]{Internal Music\\{(In-domain)}}}
        & AudioSR      & 1.84 & 1.97 & 2.79 & 1.71 & 1.89 & 2.94 & 1.56 & 1.80 & 3.08 & 1.35 & 1.77 & 3.57 \\
        & FlashSR      & 1.56 & 1.71 & 2.92 & 1.45 & 1.66 & 3.04 & 1.34 & 1.62 & 3.18 & 1.14 & 1.57 & 3.63 \\
        & FlowHigh     & 3.57 & 3.93 & 1.68 & 2.85 & 3.31 & 1.76 & 2.34 & 2.89 & 2.05 & 1.43 & 2.02 & 3.27 \\
        & LatentFlowSR & \textbf{1.23} & \textbf{1.36} & \textbf{2.97} & \textbf{1.11} & \textbf{1.29} & \textbf{3.20} & \textbf{1.04} & \textbf{1.29} & \textbf{3.53} & \textbf{0.89} & \textbf{1.29} & \textbf{4.21} \\
        \midrule

        \multirow{4}{*}{\shortstack[c]{MUSDB18-HQ\\{(Out-of-domain)}}}
        & AudioSR      & 2.06 & 2.24 & 2.58 & 1.83 & 2.08 & 2.89 & 1.63 & 1.95 & 3.13 & 1.40 & 1.89 & 3.60 \\
        & FlashSR      & 1.71 & 1.88 & 2.79 & 1.50 & 1.73 & 3.17 & 1.36 & 1.67 & 3.40 & 1.19 & 1.66 & 3.50 \\
        & FlowHigh     & 3.35 & 3.64 & 1.67 & 2.91 & 3.27 & 1.69 & 2.46 & 2.89 & 1.87 & 1.77 & 2.24 & 2.77 \\
        & LatentFlowSR & \textbf{1.41} & \textbf{1.55} & \textbf{2.81} & \textbf{1.23} & \textbf{1.43} & \textbf{3.19} & \textbf{1.15} & \textbf{1.43} & \textbf{3.46} & \textbf{0.97} & \textbf{1.41} & \textbf{4.14} \\
        \bottomrule
    \end{tabular}}
\end{table*}

\section{Experimental Setup}
\label{sec:exp}
\subsection{Datasets}
To comprehensively evaluate the super-resolution capability of the proposed LatentFlowSR across different types of audio content, we conducted experiments on three audio scenarios with the distinct time–frequency structures and high-frequency detail distributions: speech, sound effects, and music. 
This setup enabled a thorough assessment of the model’s reconstruction ability and generalization performance under diverse acoustic conditions.
\begin{itemize}[leftmargin=*]
\item {}{\textbf{Speech Dataset}:}
We used the VCTK dataset \cite{yamagishi2019cstr} for training and in-domain evaluation. 
Following previous work \cite{ai2024apcodec,lu2024towards}, the data from 108 speakers were split into a training set comprising 100 speakers (40,936 samples) and a test set comprising 8 speakers (2,937 samples).

\item {}{\textbf{Sound Effects Dataset}:}
We used the FSD50K dataset \cite{fonseca2021fsd50k} as the training set (containing 51,197 samples) to provide diverse samples of environmental sounds and event-based sound effects. 
The evaluation was then conducted on the out-of-domain ESC-50 dataset \cite{piczak2015esc} (containing 2,000 samples) to assess its generalization ability under a cross-dataset setting, thereby providing a more rigorous evaluation of its super-resolution performance on previously unseen sound effect distributions.

\item {}{\textbf{Music Dataset}:}
We used an internal dataset\footnote{Metadata for the internal dataset is available at:\url{https://github.com/fliu215/LatentFlowSR}.} containing songs in multiple languages and covering a variety of musical styles and recording conditions. 
The training set consisted of 3,962 songs, while the in-domain test set contained 87 songs. 
In addition, we also evaluated on the out-of-domain test set of the public MUSDB18-HQ dataset \cite{MUSDB18HQ} (containing 50 songs), which further helps verify the generalization ability across different types of musical content.
\end{itemize}

In the experiments\footnote{Audio samples are available at: \url{https://latentflowsr-demo.pages.dev/}.}, all audio signals are resampled to 44.1 kHz and treated as the high-resolution audio. To construct low-resolution inputs with different levels of degradation, the high-resolution audio is downsampled to 8 kHz, 12 kHz, 16 kHz, and 24 kHz, and then uniformly upsampled back to 44.1 kHz. In this way, four audio super-resolution tasks with different bandwidth limitations are constructed for each audio type.

\subsection{Implementation Details}

For the noise-robust autoencoder used in LatentFlowSR, its encoder adopted a four-stage downsampling architecture, with downsampling factors of 2, 4, 8, and 8 at each stage, respectively. 
Its decoder followed a symmetric four-stage upsampling structure, with upsampling factors of 8, 8, 4, and 2. 
Within each residual block, dilated convolutions with a kernel size of 7 (dilation rates of 1, 3, and 9) were used, along with a $1\times1$ convolution. 
All convolutional layers employed weight normalization. 
The resulting latent representation had 64 channels (i.e., $C=64$).

For the velocity field estimation network in the CFM mechanism of LatentFlowSR, its main branch of each ResNet block used one-dimensional convolutions with a kernel size of 3, while the residual branch employed a $1\times1$ convolution for projection. 
Each Transformer employed 2 attention heads.
The downsampling convolutions used a kernel size of 3 with a stride of 2, and the upsampling transposed convolutions used a kernel size of 4 with a stride of 2. 
Finally, the projection block performed feature mapping using a one-dimensional convolution with a kernel size of 3 followed by a $1\times1$ convolution.
During inference, an Euler-based ODE solver with 1 (i.e., $N=1$) iterations is used to discretely solve the ODE, enabling efficient audio super-resolution in the latent space.

We adopted a two-stage training strategy for LatentFlowSR. 
First, the noise-robust autoencoder was trained and then frozen to produce latent training data, based on which the velocity field estimation network in the CFM mechanism was subsequently trained in the learned latent space. 
Both of them were optimized using AdamW optimizer, with $\beta_{1}=0.8,\beta_{2}=0.99$, and were trained for 400k and 2M steps, respectively. 
The learning rate was set initially to 0.0003 and scheduled to decay with a factor of 0.999 at every epoch.

\subsection{Baseline Methods}
We compared the proposed LatentFlowSR with several baseline methods.
In the speech scenario, the selected baselines included NU-Wave \cite{lee2021nu}, NU-Wave2 \cite{han2022nu}, UDM+ \cite{yu2023conditioning}, AudioSR \cite{liu2024audiosr}, FlashSR \cite{im2025flashsr}, FlowHigh \cite{yun2025flowhigh}, mdctGAN \cite{shuai2023mdctgan} and AP-BWE \cite{lu2024towards}.
Among them, NU-Wave, NU-Wave2, UDM+, AudioSR, and FlashSR are diffusion-based methods, with the former two operating in the waveform domain and the latter three in the frequency domain. 
FlowHigh is a CFM-based method that performs modeling in the frequency domain, while mdctGAN and AP-BWE are discriminative methods in the same domain. 
In the sound effects and music scenarios, we compared only with methods specifically designed for general audio super-resolution (i.e., AudioSR, FlashSR, and FlowHigh), since the other approaches mainly focus on the speech domain.
For NU-Wave2, UDM+, mdctGAN, AP-BWE, and FlowHigh, we retrained their official implementations on our datasets. For NU-Wave, AudioSR, and FlashSR, we directly used the official checkpoints provided by the authors, and the output was downsampled to 44.1 kHz.
Notably, NU-Wave did not conduct the experiment at source sampling rates of 8 kHz and 12 kHz, so we excluded these results from our analysis.

\subsection{Quality Evaluation Metrics}
We adopted both objective and subjective metrics to evaluate LatentFlowSR and the baseline methods.
For objective evaluation, we used several widely used metrics in audio super-resolution \cite{lee2021nu,han2022nu,yu2023conditioning,liu2024audiosr,im2025flashsr,yun2025flowhigh,lu2024towards,li2025audio}, including log-spectral distance (LSD), high-frequency log-spectral distance (LSD-HF), and virtual speech quality objective listener (ViSQOL) \cite{chinen2020visqol,hines2015visqolaudio}. 
LSD measures the log-spectral distance between the reconstructed and target high-resolution audio over the entire frequency band, whereas LSD-HF evaluates it only within the high-frequency subband, thereby providing a more focused assessment of the model’s ability to recover high-frequency details.
ViSQOL is an objective perceptual evaluation metric that further reflects quality changes in the reconstructed high-resolution audio in a way that aligns with human auditory perception.

For subjective evaluation, we conduct mean opinion score (MOS) tests to assess the naturalness of the reconstructed high-resolution audio.
For simplicity, we conduct MOS evaluations only under the most challenging 8 kHz to 44.1 kHz super-resolution setting.
In each MOS test, twenty samples from the test set were evaluated by at least 30 listeners on the crowd-sourcing platform Amazon Mechanical Turk. For each sample, listeners were asked to rate a naturalness score between 1 and 5 with an interval of 0.5. All the MOS results were reported with 95\% confidence intervals.

\begin{table}[t]
    \centering
    \caption{Subjective MOS results for super-resolution methods on different datasets under the 8 kHz to 44.1 kHz setting.}
    \small
    \setlength{\tabcolsep}{2pt}
    \resizebox{\columnwidth}{!}{
    \begin{tabular}{ccccc}
    \hline
        Method & VCTK & ESC-50 & Internal Music & MUSDB18-HQ \\ \hline
        Low-Resolution & 2.07 $\pm$ 0.05 & 2.18 $\pm$ 0.05 & 2.53 $\pm$ 0.05 & 2.67 $\pm$ 0.05 \\ \hline
        NU-Wave2 & 3.48 $\pm$ 0.05 & - & - & - \\
        UDM+ & 3.50 $\pm$ 0.06 & - & - & - \\
        mdctGAN & 3.53 $\pm$ 0.06 & - & - & - \\
        AP-BWE & 4.01 $\pm$ 0.05 & - & - & - \\
        AudioSR & 3.58 $\pm$ 0.05 & 3.55 $\pm$ 0.06 & 3.35 $\pm$ 0.06 & 3.42 $\pm$ 0.05 \\
        FlashSR & 3.38 $\pm$ 0.06 & 3.76 $\pm$ 0.06 & 3.78 $\pm$ 0.05 & 3.95 $\pm$ 0.05 \\
        FlowHigh & 3.93 $\pm$ 0.06 & 3.18 $\pm$ 0.05 & 3.12 $\pm$ 0.05 & 3.11 $\pm$ 0.05 \\
        LatentFlowSR & \textbf{4.03 $\pm$ 0.05} & \textbf{3.97 $\pm$ 0.05} & \textbf{4.02 $\pm$ 0.05} & \textbf{3.99 $\pm$ 0.05} \\ \hline
        Ground Truth & 4.35 $\pm$ 0.05 & 4.87 $\pm$ 0.05 & 4.88 $\pm$ 0.05 & 4.80 $\pm$ 0.05 \\ \hline
    \end{tabular}}
    \label{tab:mos}
\end{table}

\section{Results and Analysis}
\label{sec:res}
\subsection{Main Results}

The objective and subjective results on the test sets of the four audio types are presented in Tables \ref{tab:all_datasets} and \ref{tab:mos}, respectively.
For in-domain speech evaluation, in the most challenging 8 kHz to 44.1 kHz task, the proposed LatentFlowSR obtained the best results on LSD-HF and ViSQOL, and matched AP-BWE on LSD. 
As the input sampling rate increased, LatentFlowSR continued to outperform the baseline methods across all three metrics, demonstrating its effectiveness under both severe and mild degradation conditions. 
These results suggest that the proposed latent-space CFM mechanism can more effectively characterize the mapping from low-resolution speech to high-resolution speech, resulting in improved spectral reconstruction and perceptual quality.
In addition, the subjective evaluation results indicate that LatentFlowSR achieved a higher MOS score than all baseline methods on the VCTK dataset, further validating its effectiveness in speech super-resolution.
Consistent with the objective results on the in-domain speech test set, LatentFlowSR also achieved the best performance across all metrics under all four super-resolution settings on the out-of-domain ESC-50 sound effects test set. 
This shows that the proposed method not only generalized well beyond speech signals, but also maintained strong robustness when evaluated on out-of-domain sound effect data, highlighting its superior generalization ability across different audio types and domains. 
The MOS results on ESC-50 further support this observation: LatentFlowSR obtained the highest scores, improving subjective ratings by more than 80\% over the low-resolution input and by approximately 5\%–6\% over the strongest baseline, indicating a clearer and more natural listening experience.


Compared with speech and sound effects, music super-resolution is generally more challenging due to its more complex harmonic structures, richer spectral details, wider frequency range, and longer temporal dependencies, which require stronger modeling capacity for both global structure and fine-grained high-frequency recovery. 
Despite these challenges, as shown in Table \ref{tab:all_datasets}, LatentFlowSR still delivered the best objective performance across all four super-resolution settings on both the in-domain internal music and out-of-domain MUSDB18-HQ test sets.
On the in-domain set, it improved LSD and LSD-HF by approximately 0.3 points over the strongest baseline FlashSR and achieved an average relative improvement of about 5\% in ViSQOL. 
On MUSDB18-HQ, LatentFlowSR consistently remained the best-performing method, demonstrating strong generalization across different music domains and degradation levels. 
Notably, although AudioSR and FlashSR have been trained on MUSDB18-HQ, they still failed to outperform LatentFlowSR. 
FlowHigh performed substantially worse on the more challenging music datasets than on the speech and sound effect datasets.
Consistent with the objective results, LatentFlowSR also achieved the best subjective performance on two music test sets under the 8 kHz to 44.1 kHz setting.
The above results provide compelling evidence that LatentFlowSR delivers high-quality reconstruction and strong generalization across diverse audio types, unseen data, and a wide range of super-resolution settings.

\begin{table}[t]
    \centering
    \caption{Complexity evaluation results between LatentFlowSR and baseline methods.}
    \begin{tabular}{cccc}
    \hline
        Method & Inference Steps & Para. $\downarrow$ & FLOPs $\downarrow$\\ \hline
        AudioSR & 50 & 258.20 M & 1213 G \\
        FlashSR & 1 & 258.20 M & 12.13 G \\
        FlowHigh & 1 & 49.40 M & 25.30 G \\
        LatentFlowSR & 1 & \textbf{10.94 M} & \textbf{0.96 G} \\ \hline
    \end{tabular}
    \label{tab:effi}
\end{table}

\subsection{Complexity Analysis of Super-Resolution}
We further analyzed the model and computational complexity of different super-resolution methods using parameter count (Para.) and floating-point operations (FLOPs) required to generate one second of audio, respectively. 
For clarity, we consider only methods specifically designed for audio super-resolution (i.e., AudioSR, FlashSR and FlowHigh), all of which are based on either diffusion models or CFM, and limit the analysis to the efficiency of super-resolution modeling in their respective representation spaces. 
The results are shown in Table \ref{tab:effi}. 
Overall, LatentFlowSR achieved the lowest parameter count and FLOPs with only one inference step, demonstrating clear complexity and efficiency advantages.
AudioSR required as many as 50 diffusion-based sampling steps and incurred over 200M parameters and more than 1000G FLOPs. 
Built upon AudioSR, FlashSR used distillation to reduce the inference sampling steps to 1 and disable classifier-free guidance (CFG), thereby reducing FLOPs by about an order of magnitude, but without lowering model complexity.
With the same one-step inference setting, LatentFlowSR incurs only about 8\% of the FLOPs of FlashSR, which may be attributed to the fact that its parameter count was only 4.2\% of that of FlashSR.
Compared with FlowHigh, a CFM-based method operating in the mel-spectrogram domain, LatentFlowSR also achieved clear reductions in both parameter count and FLOPs, indicating that performing CFM-based super-resolution in the latent space is considerably more efficient. 
These results suggest that latent-space modeling can effectively reduce the difficulty of audio super-resolution, thus enabling the adoption of a more lightweight architecture with lower computational complexity and improved efficiency.

\subsection{Analysis of Autoencoder’s Reconstruction Quality and Noise Robustness}
\begin{table}[t]
    \centering
    \caption{Reconstruction quality evaluation results between proposed noise-robust autoencoder and other methods.}
    \begin{tabular}{cccc}
    \hline
        Method & Type & LSD $\downarrow$ & ViSQOL $\uparrow$ \\ \hline
        DAC & VQ-VAE & 0.93 & 4.13 \\
        BigVGAN & vocoder & 1.14 & 3.17 \\
        HiFi-GAN & vocoder & 1.73 & 3.09 \\
        Plain AE & autoencoder & 0.94 & 4.11 \\
        NRAE & autoencoder & \textbf{0.92} & \textbf{4.31} \\ \hline
    \end{tabular}
    \label{tab:quality}
\end{table}

\begin{table}[t]
    \centering
    \caption{Noise robustness evaluation results of the proposed noise-robust autoencoder and the plain autoencoder.}
    \label{tab:nr}
    \setlength{\tabcolsep}{4pt}
    \small
    \resizebox{\columnwidth}{!}{
    \begin{tabular}{ccccccc}
        \toprule
        \multirow{2}{*}{Method} & \multicolumn{2}{c}{1$\times$ Latent Noise} 
        & \multicolumn{2}{c}{2$\times$ Latent Noise} 
        & \multicolumn{2}{c}{3$\times$ Latent Noise} \\
        \cmidrule(lr){2-3} \cmidrule(lr){4-5} \cmidrule(lr){6-7}
        & LSD $\downarrow$ & ViSQOL $\uparrow$
        & LSD $\downarrow$ & ViSQOL $\uparrow$
        & LSD $\downarrow$ & ViSQOL $\uparrow$ \\
        \midrule
        Plain AE & 3.42 & 2.36 & 3.72 & 2.38 & 3.83 & 2.30 \\
        NRAE & \textbf{0.92} & \textbf{4.22} & \textbf{0.93} & \textbf{4.18} & \textbf{0.94} & \textbf{4.11} \\
        \bottomrule
    \end{tabular}}
\end{table}
The proposed noise-robust autoencoder (denoted by NRAE) is a key component of LatentFlowSR, and its reconstruction quality and noise robustness directly affect the performance of latent-space super-resolution. 
To evaluate its effectiveness, we compared the noise-robust autoencoder on the challenging internal music dataset with other analysis–reconstruction methods, including DAC \cite{kumar2023high}, a vector-quantized variational autoencoder (VQ-VAE) based method; BigVGAN \cite{lee2023bigvgan} and HiFi-GAN \cite{kong2020hifi}, two vocoder-based methods; and a plain autoencoder (denoted by Plain AE), i.e., training without latent noise injection. 
Note that this experiment only evaluates the analysis–reconstruction performance of each method, i.e., encoding high-resolution audio into intermediate representations (i.e., quantized result, mel-spectrogram or latent representation) and then reconstructing it, without involving the super-resolution process.
Therefore, we adopted only LSD and ViSQOL as evaluation metrics. 
As shown in Table \ref{tab:quality}, the noise-robust autoencoder achieved the best reconstruction performance among all compared methods. 
Compared with the plain autoencoder, the noise-robust variant further improved ViSQOL, indicating that latent noise injection helped preserve reconstruction fidelity.
The noise-robust autoencoder also outperformed DAC and two vocoder-based methods, suggesting that continuous latent representations are more suitable for high-fidelity audio reconstruction compared with discrete representations and mel-spectrograms.

We further analyzed the noise robustness of the proposed noise-robust autoencoder. 
During inference, Gaussian noise scaled by factors of 1, 2, and 3 was injected into the latent representations of the noise-robust and plain autoencoders, respectively.
The reconstruction quality results are shown in Table \ref{tab:nr}. 
As the noise level increased, the performance of plain autoencoder degraded significantly, whereas the noise-robust autoencoder remained relatively stable. 
These results demonstrate that the noise-robust autoencoder possessed strong robustness to latent perturbations, thereby providing a more stable latent space for subsequent super-resolution.

\subsection{Experimental Analysis of Inference Step Count}

As shown in Table \ref{tab:step}, LatentFlowSR achieves the best performance when using only one inference step, yielding the lowest LSD and LSD-HF as well as the highest ViSQOL. As the number of inference steps increases from 1 to 8, all metrics show a slight but consistent degradation. Based on this observation, we choose one-step inference as the default setting, since it not only gives the best reconstruction quality but also provides the highest inference efficiency. This result suggests that super-resolution in the proposed latent space is sufficiently simple and well-structured, such that the target high-resolution representation can be recovered effectively with only a single update. In contrast, using more inference steps does not bring further benefits and may even accumulate prediction and integration errors along the trajectory. In particular, since the velocity field is learned from randomly sampled intermediate states during training, multi-step rollout at inference may introduce a mismatch, where later states deviate from the latent distribution seen during training. As a result, repeated updates can gradually amplify small errors, leading to slight performance degradation.

\begin{table}[t]
    \centering
    \caption{Experimental results of LatentFlowSR with different inference steps in 8 kHz to 44.1 kHz setting on the internal music dataset.}
    \begin{tabular}{cccc}
    \hline
        Inference Steps & LSD $\downarrow$ & LSD-HF $\downarrow$ & ViSQOL $\uparrow$ \\ \hline
        1 & \textbf{1.231} & \textbf{1.356} & \textbf{2.969} \\
        2 & 1.236 & 1.361 & 2.954 \\
        4 & 1.237 & 1.363 & 2.953 \\
        8 & 1.239 & 1.365 & 2.946 \\ \hline
    \end{tabular}
    \label{tab:step}
\end{table}

\subsection{Experiment Analysis of Low-Frequency Replacement}
As shown in Table \ref{tab:lfr}, introducing low-frequency replacement (LFR) consistently improves the performance of LatentFlowSR on the 8 kHz to 44.1 kHz super-resolution task for the internal music dataset. Specifically, removing LFR leads to a degradation from 1.23 to 1.27 in LSD and from 2.97 to 2.94 in ViSQOL. This indicates that LFR helps preserve the reliable low-frequency information provided by the input, allowing the model to focus more effectively on restoring the missing high-frequency content. As a result, the proposed method achieves lower spectral distortion and better perceptual quality. These results confirm that explicitly constraining the low-frequency part is beneficial for super-resolution.

\begin{table}[t]
    \centering
    \caption{Effect of low-frequency replacement (LFR) on LatentFlowSR in the 8 kHz to 44.1 kHz setting on the internal music dataset.}
    \begin{tabular}{cccc}
    \hline
        Method & LSD $\downarrow$ & ViSQOL $\uparrow$ \\ \hline
        LatentFlowSR & \textbf{1.23} & \textbf{2.97} \\
        LatentFlowSR w/o LFR & 1.27 & 2.94 \\ \hline
    \end{tabular}
    \label{tab:lfr}
\end{table}

\subsection{Ablation Studies}
\begin{table}[t]
    \centering
    \caption{Ablation results of LatentFlowSR under the 8 kHz to 44.1 kHz setting.}
    \begin{tabular}{cccc}
    \hline
        Method & LSD $\downarrow$ & LSD-HF $\downarrow$ & ViSQOL $\uparrow$ \\ \hline
        LatentFlowSR & \textbf{1.23} & \textbf{1.36} & \textbf{2.97} \\
        LatentFlowSR w/o NR & 1.30 & 1.43 & 2.89 \\
        LatentFlowSR w/ Mel & 1.32 & 1.42 & 2.83 \\
        LatentFlowSR w/ DAC & 1.29 & 1.42 & 2.91 \\ \hline
    \end{tabular}
    \label{tab:ablation}
\end{table}
Finally, we conducted ablation studies of LatentFlowSR under the most challenging 8 kHz to 44.1 kHz setting on the internal music dataset, with results shown in Table \ref{tab:ablation}. 
Specifically, to examine the effect of noise-robust training, we replaced the proposed noise-robust autoencoder with plain one and performed super-resolution in its latent space, denoted as LatentFlowSR w/o NR. 
We also replaced the proposed latent modeling space with the mel-spectrogram space and the discrete latent space constructed by DAC, denoted as LatentFlowSR w/ Mel and LatentFlowSR w/ DAC, respectively.
As shown in Table \ref{tab:ablation}, removing noise-robust training led to consistent degradation across all metrics. 
This indicates that when the predicted latent representation deviates from the ground-truth latent representation, the noise-robust latent space can better compensate for this mismatch, thereby improving the quality of reconstructed audio and the recovery of high-frequency details. 
In addition, replacing the proposed latent space with either the mel-spectrogram space or the discrete latent space also caused clear performance drops. 
These results suggest that the proposed noise-robust latent space in LatentFlowSR is better suited for audio super-resolution, owing to its stronger capacity to support high-quality latent modeling and more accurate recovery of high-frequency details.

\section{Conclusion}
\label{sec:con}
In this paper, we propose LatentFlowSR, an audio super-resolution method based on CFM mechanism in the latent space. 
Unlike approaches that directly perform high-dimensional generation in the waveform or spectral domain, our method first employs a noise-robust autoencoder to map audio into a continuous latent space. 
Conditioned on the low-resolution latent representation, a CFM mechanism, whose velocity field is parameterized by a U-Net-style hierarchical architecture, is then used to generate high-resolution latent representations from a Gaussian noise prior.
During inference, only a one-step ODE solver is required to obtain the high-resolution latent representation, which is finally decoded by the pretrained autoencoder to reconstruct the high-resolution audio. 
Experimental results strongly demonstrate the effectiveness of latent-space modeling for audio super-resolution.
In future work, we plan to further improve super-resolution performance, explore a wider range of audio types, and investigate approaches to achieve even higher computational efficiency.

\bibliographystyle{IEEEtran}
\bibliography{mybib}
\end{document}